\documentclass{optica-article}

\journal{opticajournal} 

\articletype{Research Article}

\usepackage{lineno}

\usepackage{tikz}

\graphicspath{{figures/}}

\definecolor{lightblue}{rgb}{0.3569,0.502,0.7216}

\def\linkcolor{lightblue}

\usepackage{hyperref}
\hypersetup{
    colorlinks=true,
    linkcolor=\linkcolor, 
    urlcolor=\linkcolor,
    citecolor=\linkcolor
}

\DeclareUnicodeCharacter{3000}{\textcolor{red}{BAD!!}}

\DeclareUnicodeCharacter{2009}{\textcolor{red}{BAD!!}}

\DeclareUnicodeCharacter{03A3}{$\Sigma$}

\DeclareUnicodeCharacter{03A0}{$\Pi$}

\DeclareUnicodeCharacter{03BC}{$\mu$}

\DeclareUnicodeCharacter{03B2}{$\beta$}

\newcommand{\rfig}[1]{Fig.~\ref{#1}}
\newcommand{\rsec}[1]{Section~\ref{#1}}
\newcommand{\rtab}[1]{Table~\ref{#1}}

\newcommand{\company}[1]{{\it #1}}

\usetikzlibrary{patterns,arrows,calc,decorations.pathmorphing,angles,angles,quotes,arrows.meta,bending,arrows.meta}

\tikzset{>=latex} 

\newcommand{\change}[1]{{\color{black}#1}}

\usepackage{verbatim}

\begin{document}

\title{Low Duty Cycle Pulsed UV Technique for Spectroscopy of Aluminum Monochloride}

\author{
Li-Ren Liu,\authormark{1}
Brian K.~Kendrick,\authormark{2}
and 
Boerge Hemmerling\authormark{1,*}
}

\address{
\authormark{1}Department of Physics and Astronomy, University of California, Riverside, USA\\
\authormark{2}Theoretical Division (T-1, MS B221), Los Alamos National Laboratory, Los Alamos, New Mexico 87545, USA
}

\email{\authormark{*}boergeh@ucr.edu}

\homepage{http://molecules.ucr.edu}

\begin{abstract*} 
We present a novel technique to minimize UV-induced damage in experiments that employ second-harmonic generation cavities. The principle of our approach is to reduce the duty cycle of the UV light as much as possible to prolong the lifetime of the used optics. The low duty cycle is achieved by ramping the cavity into resonance for a short time during the experimental cycle when the light is used and tuning it to an off-resonant state otherwise. The necessary fast ramp and length-stabilization control of the cavity is implemented with the FPGA-based STEMlab platform. We demonstrate the utility of this method by measuring the isotope shift of the electronic transition ($X^1\Sigma \leftarrow A^1\Pi$) in AlCl at 261.5\,nm in a pulsed molecular beam experiment.
\end{abstract*}

\section{Introduction}

The invention of the laser led to a myriad of applications in science and industry \cite{Duarte2016}.
In fundamental research, the laser enabled applications ranging from precision spectroscopy and tests of fundamental theories \cite{Safronova2018}, cooling and trapping of atoms and molecules \cite{Metcalf1999,McCarron2018a}, to cold controlled chemistry \cite{Krems2008,Carr2009,Krems2019}.
This type of research often uses narrow-band continuous-wave (CW) lasers to provide a high frequency resolution. Many of these applications also require laser light in the deep ultraviolet (UV) to cover optical transitions in atoms and molecules for carrying out precision spectroscopy and laser cooling \cite{Metcalf1999}. Specific examples include atomic cadmium with a laser cooling transition at 229\,nm \cite{Brickman2007,Kaneda2016,Tinsley2021} and atomic mercury at 254\,nm \cite{Villwock2011,Zhang2021c}.

While solid-state, robust and tunable CW laser systems exist for a wide range of wavelengths in the visible and infrared, at the same time, developing such systems for the deep ultraviolet range is technically very challenging, partly because deep-UV radiation damages and degrades the involved materials.
Nevertheless, recent efforts were able to produce UV emitting diodes at 271.8\,nm \cite{Zhang2019} and ongoing research efforts keep on pushing the development of UV laser technology with a focus on AlGaN-based edge emitters \cite{Amano2020}.

In addition to a specific wavelength in the deep-UV, many of the interesting applications require the use of high UV laser power.
For instance, driving dipole-forbidden transitions, as is done for spectroscopy of hydrogen \cite{Parthey2011,Ahmadi2018,Fleurbaey2018,Grinin2020}, muonium \cite{Crivelli2018} and xenon \cite{Altiere2018}, requires deep-UV laser light at the level of Watts.
Another example involves the laser cooling of molecules \cite{Rosa2004,McCarron2018a,Tarbutt2019,Chae2023}. Two of the promising candidates, AlCl \cite{Daniel2021,Daniel2023} and AlF \cite{Hofsass2021}, have laser cooling transitions at 261\,nm and 228\,nm, respectively. These molecules potentially provide high capture velocities for magneto-optical traps of up to 30--40\,m/s \cite{Daniel2023}. To achieve this level, though, high intensities are required to saturate the cooling transitions.
At the same time, this requirement is in conflict with the desire to use laser beams with large beam cross sections.
Here, the aim is to provide a large overlap with a molecular beam to render the capture process as efficient as possible.
To address such experimental needs, tremendous progress has been made over the years and various laser systems at wavelengths in the deep-UV below 300\,nm with output levels ranging from 50\,mW to more than 1\,Watt, have been developed 
\cite{Mes2003b, McCarron2021, Zhadnov2023, Burkley2019, Hu2013, Sayama1997, Cooper2018, Imai2003, Asakawa2004, Huang2008, Oka2004, Masuda2001, Tinsley2021}.

A major challenge when using deep-UV lasers is the laser-induced damage that can occur in any of the involved optical components.
The intensity threshold, typically designated as laser-induced damage threshold (LIDT), at which significant damage occurs depends on the optical material itself and varies with the wavelength.
The mechanisms of these effects have been studied for typical used optical materials, e.g.~CaF$_2$ \cite{Bauer2009,Bauer2009a} and fused silica \cite{Schenker1994,Schenker1995,Negres2010}.
Among the detrimental effects causing damage is the absorption of moisture by materials that are hygroscopic, the breaking of chemical bonds, the depletion of oxygen and the contamination with hydrocarbons upon exposure \cite{Kunz2000,Heinbuch2008,Gangloff2015}.
Various methods have been explored to revert these effects that focus on keeping oxygen near the surface of the optics through oxide coatings \cite{Gangloff2015} or the submersion of the optics in an oxygen environment \cite{Cooper2018}. Moreover, fluoride-coated optics that avoid the need for the presence of oxygen have been shown to be beneficial \cite{Burkley2021}.
Non-linear crystals, which are used for higher-harmonic generation to produce deep-UV light, are often hygroscopic and suffer from UV-induced damage. Extensive studies have been carried out to characterize these damages and to find mitigation techniques for non-linear crystals, such as BaB$_2$O$_4$ (BBO) \cite{Watanabe1991,Takahashi2010,Turcicova2022} and CsLiB$_6$O$_{10}$ (CLBO) \cite{Nishioka2005,Kawamura2009,Takachiho2014,Yoshimura2015,Turcicova2022}, which are relevant for this work.
Optical fibers, another essential component in today's experiments involving lasers, have been shown to be able to withstand larger amounts of UV intensities if they are pretreated with hydrogen and irradiate with UV light \cite{Colombe2014}.

While the choice of materials and the developed specific treatments have shown promising results to prepare optical components for operating in the deep-UV regime, a material-independent method to reduce UV-degradation in the first place is to minimize high intensity UV-exposure of an optical material.
In some optical setups this is possible by avoiding laser beam foci near or inside optical materials, which consequently reduces the local UV intensity \cite{Hemmerling2011}.
This technique, however, cannot be used in setups where optical foci are a requirement.
Here, an alternative approach is to reduce the duty cycle of the UV light to a minimum instead.
This technique is most naturally applied in pulsed experiments that often exhibit times between experimental repeats where the laser is only idling. However, it can also be used to bridge dead times in continuous experiments.

In this work, we present a method to reduce the duty cycle of a second-harmonic generation (SHG) cavity for UV light by using an FPGA-based fast control to length-stabilize the cavity for a short period during the experiment only and move it off-resonant otherwise. 
We demonstrate an application of our technique by carrying out detailed spectroscopy on the isotope shift of the diatomic molecule AlCl, as described in \rsec{sec:isotope_shift}.
We note that recently a similar technique that reduces the duty cycle where the circulating power inside a SHG cavity was quickly switched ($\approx \mu$s) with an acousto-optic modulator has been developed \cite{Zhadnov2023}.
Both of these methods are particularly suitable for experiments geared towards fundamental research since the lack of readily available solid-state UV technology often requires the use of SHG cavities to produce UV light.
\change{
As a consequence of the reduced duty cycle, any optics that is used to steer or manipulate the UV laser light for the experiment experiences significantly less UV radiation exposure. This, in turn, prolongs the lifetime of the optical setups and renders experiments that use deep-UV light more cost-effective.
}

\change{
In the future, we plan to use our pulsed approach to implement a robust, high-power laser system to cool and load AlCl in a magneto-optical trap. Laser cooling of AlCl is expected to produce a large number of trapped molecules due to the high capture velocity, but requires high laser powers. 
The large trapped numbers are beneficial to many proposed experiments with cold molecules, including controlled chemistry \cite{Krems2008,Ni2010,Ye2018,Ospelkaus2010}, quantum simulation \cite{DeMille2002,Yelin2006,Yu2019,Carr2009,Micheli2006,Bao2022,Holland2022} or precision measurements \cite{Andreev2018,Cairncross2017,Kozyryev2017a,Hudson2011,Kozyryev2021,Kondov2019,ACMECollaboration2014,Fitch2021a,Cairncross2017,Yu2021,Hutzler2020,ORourke2019,Aggarwal2018,Uzan2003,DeMille2008,Chin2009,Kajita2009,Beloy2010,Jansen2014,Dapra2016,Kobayashi2019,Chupp2019}.
}


\section{Experimental Method}

Our method can be summarized as follows. We produce UV light at 261.5\,nm in a second-harmonic generation cavity by ramping the cavity into resonance for only 50\,ms during each experimental cycle, which repeats every $\approx 1$\,s. The molecules traverse and interact with the laser beam during this time window and the cavity is off-resonant till the next experimental cycle, effectively providing a duty cycle of the UV light of $\approx 5\%$.

The experiment starts with a cryogenic helium buffer-gas beam source \cite{Hutzler2012} to produce a beam of aluminum monochloride (AlCl) molecules. AlCl molecules are brought into the gas phase via short-pulsed laser ablation of a solid precursor made of a mixture of Al and KCl \cite{Lewis2021} at 532\,nm with a Nd:Yag laser. 
\change{
The molecular beam travels through our vacuum system and is subject to laser-induced time-resolved fluorescence spectroscopy $\approx 0.5$\,m downstream from the source. 
To provide a good signal-to-noise, the fluorescence light is imaged onto a relay image using a plano-convex lens (focal length of 75\,mm) and spatially filtered with an iris. 
The relay image is then focused with two plano-convex lenses (focal length of 100\,mm) onto a photo multiplier tube (\company{Hamamatsu, H10722-04}) and analyzed.
A UV bandpass filter (\company{Thorlabs, FGUV5M}) with a transmission of 70\% at 260\,nm is used to remove background scatter onto the PMT.
}
In the following, we focus the discussion on the spectroscopy laser setup, while further details on the cryogenic buffer-gas beam part of our apparatus can be found in Refs.~\cite{Daniel2021,Daniel2023}.

\begin{figure}[ht]
\begin{center}
\includegraphics[width=\linewidth]{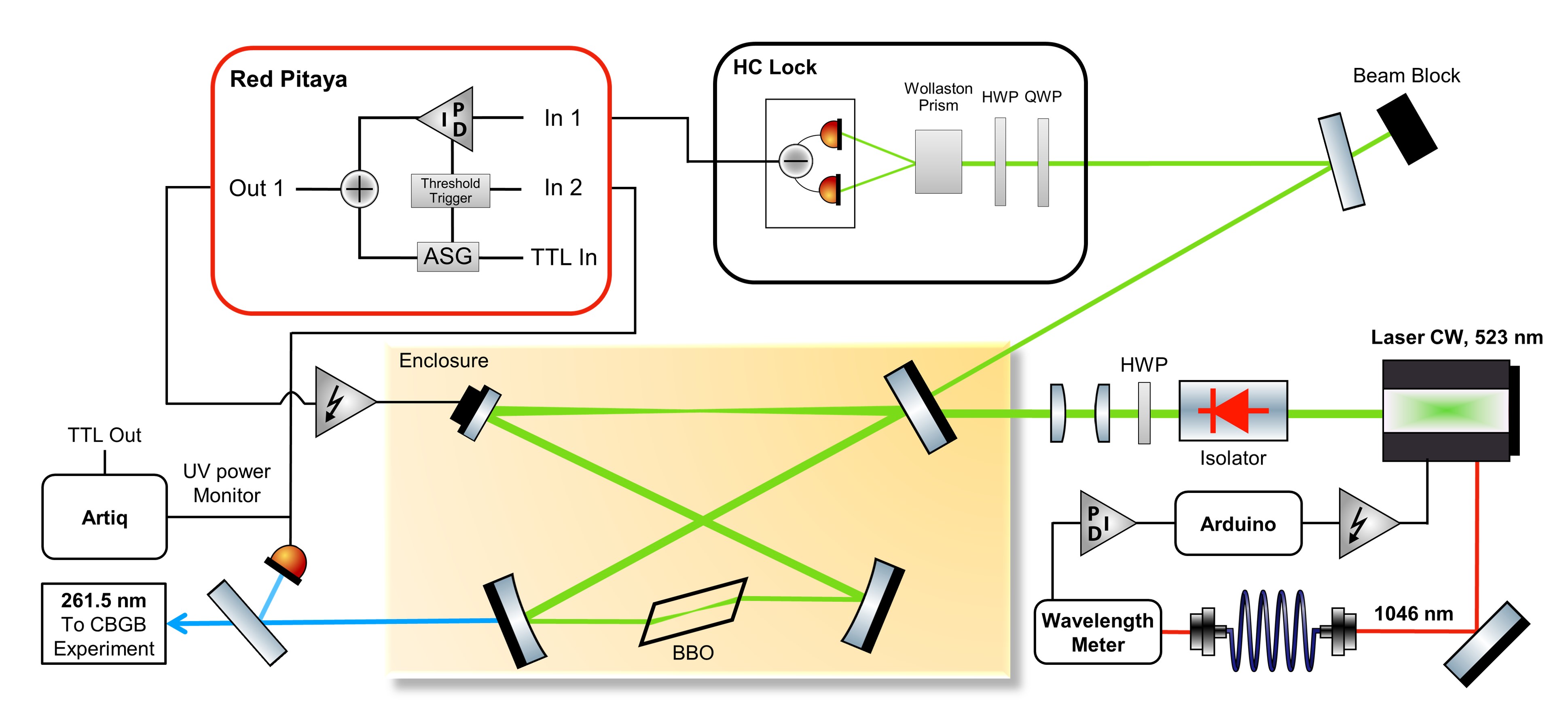}
\end{center}
\caption{
\label{fig:setup}
Schematic of the laser setup to produce laser light at 261.5\,nm with a low duty cycle.
The UV light is produced by a second-harmonic generation bow-tie cavity using a BBO crystal and the fundamental at 523\,nm.
The cavity is ramped into resonance shortly before each experimental cycle and the output power is monitored with a photodiode. Once the UV output reaches a preset threshold, the ramp is held at its current value and a PID controller is switched on to length-stabilize the cavity and keep it on resonance for a predefined time. 
}
\end{figure}

\subsection{Resonance-Triggered Stabilization Technique}
\label{sec:technique}

\begin{figure}[t]
\begin{center}
\includegraphics[width=0.55\linewidth]{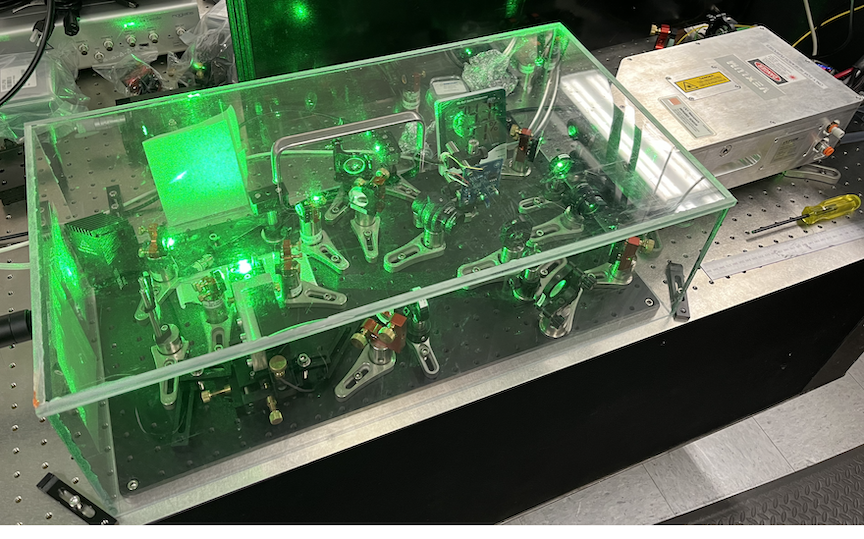}
\hspace*{0.5cm}
\includegraphics[width=0.35\linewidth]{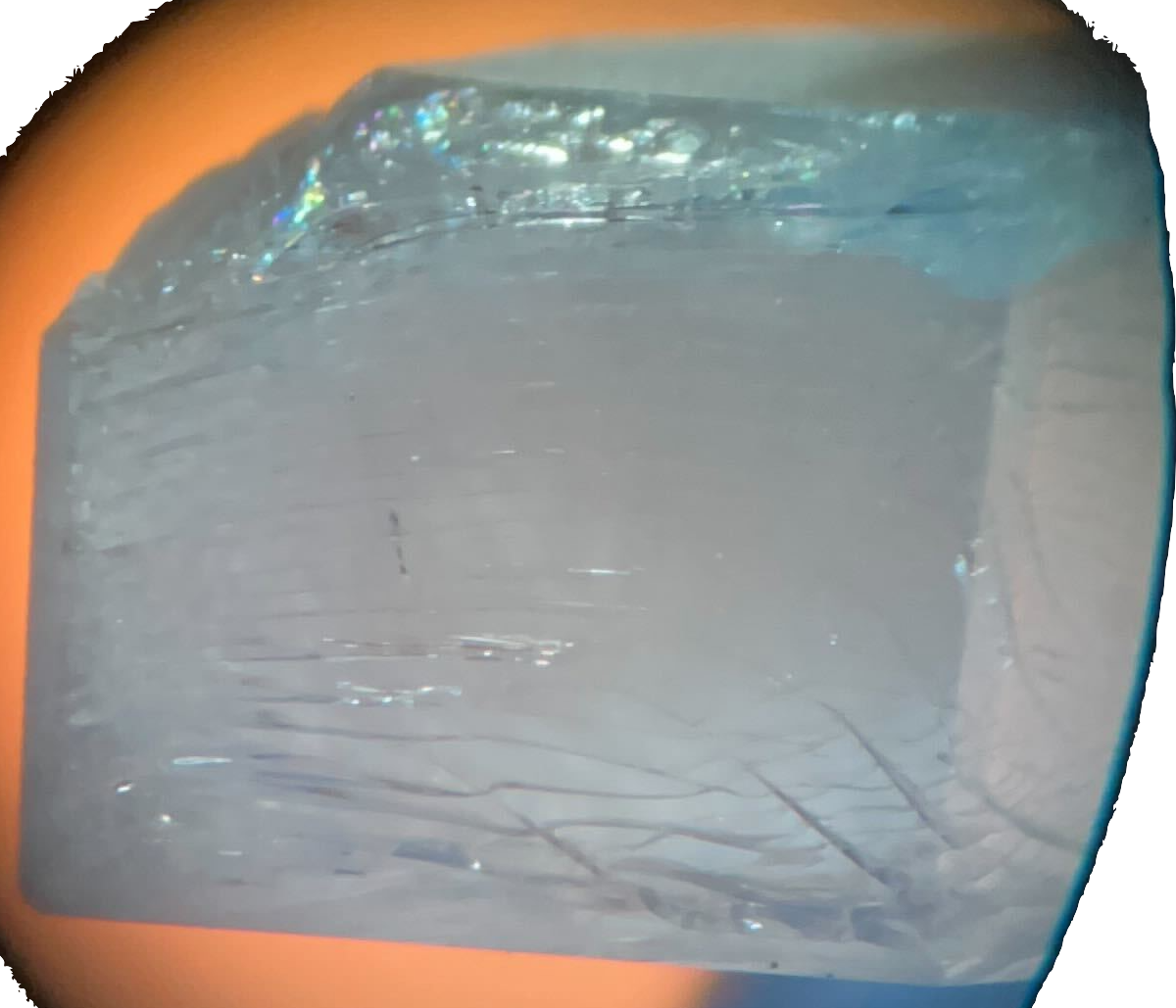}
\end{center}
\caption{
\label{fig:vexlum}
\label{fig:setup_photo}
\label{fig:crystal}
\textbf{Left:} Photo of the experimental laser setup. The enclosure houses the SHG cavity to produce 261.5\,nm laser light and is continuously flushed with dry air (moisture level below $10$\%).
\textbf{Right:} Photo taken with a microscope of the front surface of a CLBO crystal after using it in CW-mode in the SHG cavity. After a few days of operation, the crystal became very brittle and developed ruptures.
}
\end{figure}

The spectroscopy light at $261.5$\,nm is produced by frequency-doubling CW laser light at 523\,nm (\company{Vexlum, VALO SHG SF}) in a homebuilt second-harmonic generation (SHG) bow-tie cavity with a non-linear beta-barium-borate (BBO) crystal (\company{Newlight Photonics}), as shown in \rfig{fig:setup}.
\change{
The cavity consists of two curved mirrors with a radius of curvature of $50$\,mm, an input coupler with 97\% transmission at 523\,nm and a piezo-actuated mirror for length stabilization.
}
Frequency drifts of the \company{Vexlum} laser are compensated by comparing it to a wavelength meter (\company{High Finesse, WS-7}).
Specifically, a part of the light from a monitoring output of the laser at 1046\,nm is continuously measured by the wavelength meter and compared to a set frequency to generate an error signal. 
This signal is then fed into a software-based proportional-integral controller (PI) to produce a control signal, which is sent to the internal frequency-tuning piezo of the laser via the analog voltage output of a microcontroller (\company{Arduino Due}).

The absolute frequency calibration of the wavelength meter is done by regularly comparing a Doppler-free saturated absorption spectrum of rubidium to known literature values \cite{Steck2019}. This calibration step uses a separate CW Ti:Saph (\company{Coherent 899-21}) laser. The shot-to-shot variations of the measured laser frequency are a convolution of the limited resolution of the wavelength meter (specified to $\pm 2$\,MHz), the intrinsic frequency noise of the laser, the limited bandwidth of the feedback loop due to the software-based PI control and the time it takes to read out the laser frequency ($\approx 10$\,ms). Overall, these effects lead to a conservative upper limit of the frequency stability of $\pm 15$\,MHz in the UV.

The output of the 523\,nm laser is sent through an optical isolator (\company{Thorlabs, IO-5-532-HP}) and focused into the SHG cavity using mode-shaping lenses to hit the optimal target beam waist of $\approx 20\,\mu$m inside the crystal, which is determined by the Boyd Kleinman theory \cite{Kleinman1966, Boyd1968}.
For the SHG process in the BBO at this wavelength, critical type I phase matching at room temperature is achieved by using a crystal that is cut at an angle of $\theta = 48.9$\,deg.
The crystal facets are cut at the Brewster angle for the fundamental wavelength ($\theta \approx 59.2$\,deg) to minimize reflection losses at each round trip in the cavity. 
\change{
The cavity is enclosed in a box made of acrylic with dimensions 62 x 33 x 19 cm
, as shown in \rfig{fig:setup_photo}, and the box is continuously flushed with compressed air from the building supply lines with an airflow of $\approx 1$\,liter/min.
}
To minimize contamination, an air filter that blocks particulates larger than 0.01\,\textmu m (\company{Parker, 9933-11-BQ}) is installed inline. After flushing the enclosure for $\approx 10$\,min, a humidity level of $\le 10\%$ and a clean environment is achieved around the cavity.
The measured finesse of the cavity is approximately $140$ and the incoupling efficiency is $\approx 27\%$. For the measurements presented in this work, we typically operated with a few milliwatt UV power.
\change{
We didn't optimize these cavity parameters further due to degradation issues of the crystal.
}
Moreover, the measurements presented here only needed low UV power but required us to still use our resonance-triggered stabilization technique. For future experiments that require higher UV power, we will use a CLBO crystal for the SHG cavity. \change{More details on these points are discussed in \rsec{sec:improvements}.}

Resonant enhancement of the non-linear conversion process is achieved by length-stabilizing the SHG cavity to the fundamental wavelength using the H\"ansch-Couillaud (HC) technique \cite{Hansch1980}. The error signal created by measuring the difference of the two quadratures in the HC setup is fed into the fast 14-bit ADC input of a field-programmable gate array (FPGA)-based platform (\company{Red Pitaya, STEMLab 125-14}, clocked at 125 Msps) for further processing.
The 125-14 board has found wide applications in experimental physics and has been used and characterized in various experiments to stabilize cavities and lock lasers \cite{
Hannig2018,
Preuschoff2020, Pultinevicius2023,
Lauria2022,
Kestler2023
}. To program the FPGA, we built on and customized the modules from the open source software package Python Red Pitaya Lockbox (PyRPL) \cite{Neuhaus2024}.

As shown in \rfig{fig:setup}, the error signal is sent via the ADC input (In 1) to the PI module, which produces a control signal at the fast DAC output (Out 1) of the FPGA board. This signal is then amplified by an HV amplifier (\company{TEM Messtechnik, miniPIA}) to drive a piezoactuated cavity mirror.
The control signal has added the output of an arbitrary signal generator module (ASG), which is used to apply a voltage ramp to the piezo.
The second ADC input (In 2) monitors the UV intensity output of the cavity by reading a photodiode that picks up a small fraction of the output beam.
When the UV intensity reaches a predefined threshold, an internal threshold trigger is latched that switches on the PI control module and switches the ASG output to a hold state, maintaining its current voltage output.
This step switches off the voltage ramp and switches on the feedback loop for a predefined time interval.

\begin{figure}[ht]
\begin{center}
\includegraphics[width=\linewidth]{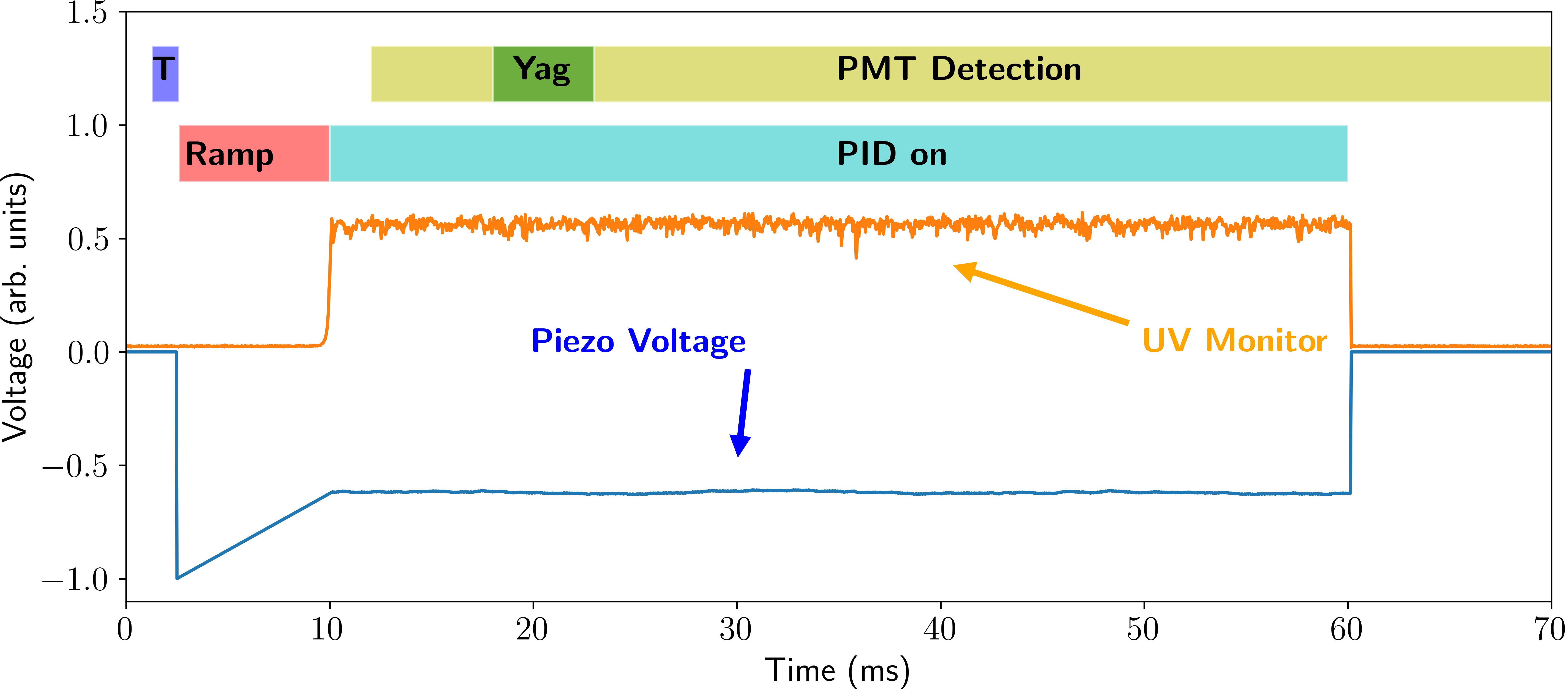}
\end{center}
\caption{
\label{fig:ramp_lock}
\label{fig:sequence}
Experimental time sequence of a single ablation shot.
At the start of each experimental cycle (Trigger T), a ramp is applied to the piezo of the SHG cavity mirror. Once the cavity becomes resonant with the fundamental light at 523\,nm and the UV output increases. This output is monitored with a photodiode and activates a trigger once a set threshold is reached. Upon the trigger, the piezo ramp is set to hold the voltage and the PI feedback for length stabilization of the cavity is switched on. After the experiment, the cavity is deliberately tuned out of resonance.
}
\end{figure}

The complete experimental sequence, shown in \rfig{fig:sequence}, is as follows: First, the main laser wavelength is tuned and stabilized to a given set point by the computer. Then, our data acquisition system Artiq \cite{Artiq2021} sends a TTL trigger (T) to the FPGA board to initiate the voltage ramp (Ramp). The FPGA switches on the cavity stabilization (PID on) as soon as it reaches the resonance condition. After a short delay, which is chosen such that the cavity resonance condition is always met by ramping over at least one free spectral range, Artiq triggers an ablation laser shot (Yag) to produce a beam of molecules.
The SHG cavity is then held on resonance for a time interval that is sufficient to let the molecules traverse the experimental apparatus ($\approx 50$\,ms) and be detected on \change{the} photomultiplier (PMT detection).
After this time, the PI control module is switched off and the piezo voltage ramp output is set to zero to intentionally tune the cavity out of resonance and eliminate the UV output. This sequence is repeated for a preset number of averages after which the laser wavelength is tuned to a new set point.
An example of the voltage ramp (solid blue curve) and the UV output intensity (solid orange curve) of a single experimental cycle is shown in \rfig{fig:sequence}.

\subsection{Measurement and Ab Initio Calculations of the Isotope Shift in AlCl}
\label{sec:isotope_shift}

\begin{figure}[t]
\begin{center}
\begin{tikzpicture}

\tikzstyle{ell}=[{Latex[length=3.3,width=2.2]}-{Latex[length=3.3,width=2.2]},line width=0.3]

    \node at (0,0) {\includegraphics[width=\linewidth]{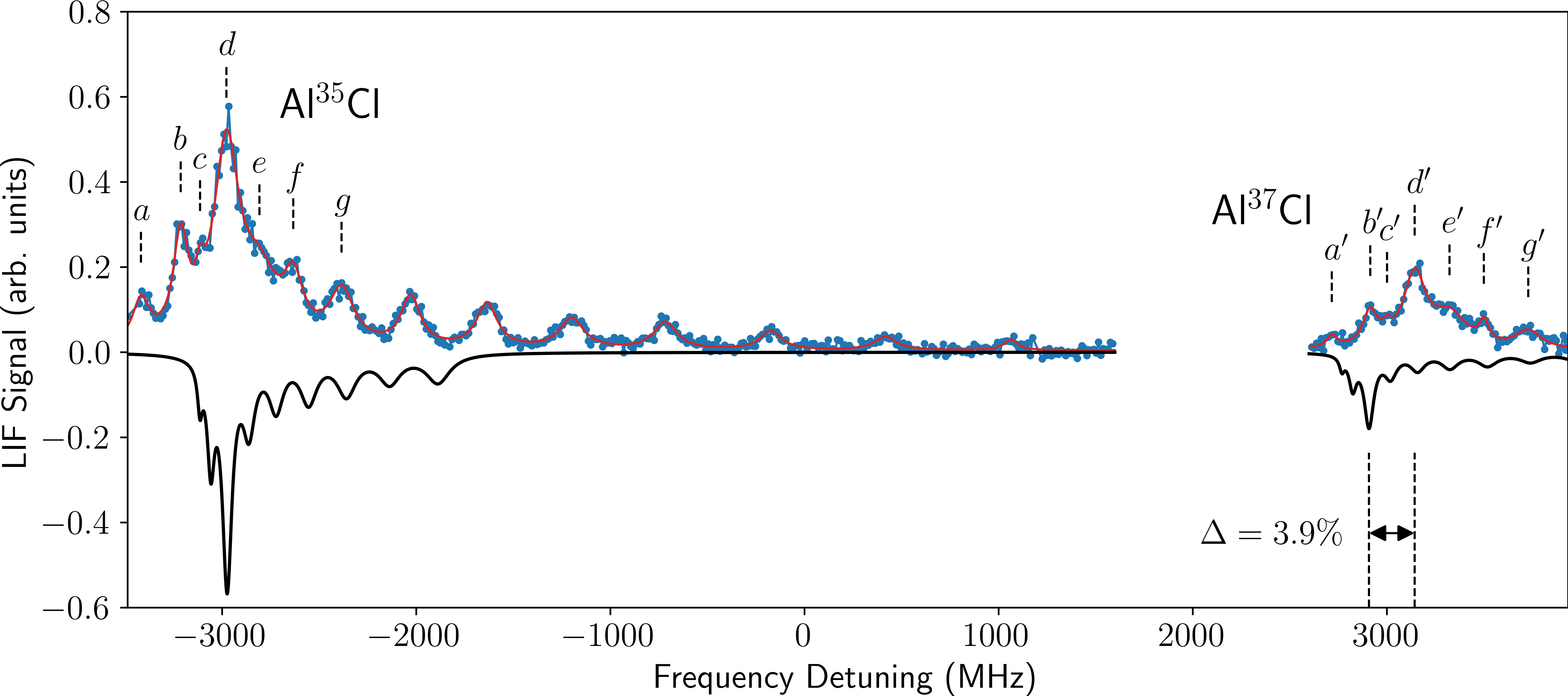}};
  	
   	\draw [draw=white, fill=white] (-4.5cm, 1.75cm) rectangle (-3.0cm, 2.5cm);
	\draw [draw=white, fill=white] (3.5cm, 1.0cm) rectangle (4.5cm, 1.5cm);
    \node at (-3.75cm, 2.0cm) {Al$^{35}$Cl};
	\node at (5.5cm, 2.0cm) {Al$^{37}$Cl};

	\begin{scope}[xshift=0.5cm]
    	\node at (0.25cm,1.7cm) {\includegraphics[width=0.45\linewidth]{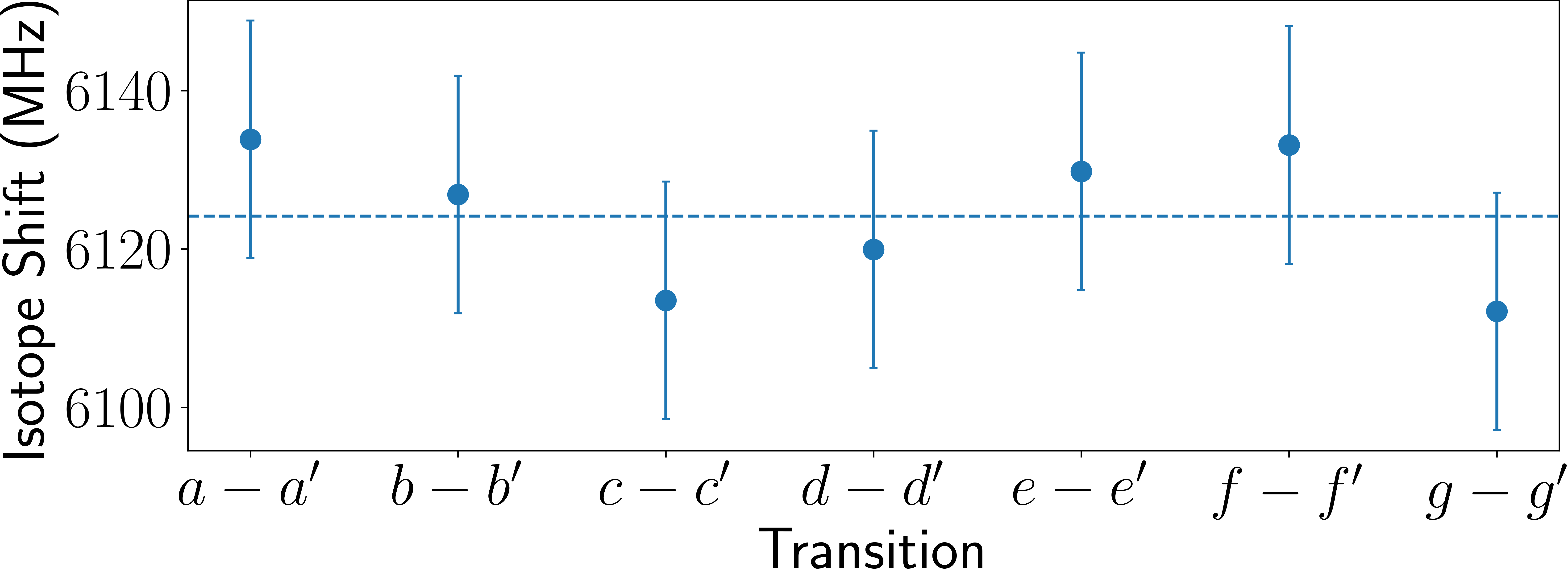}};
	    \draw [draw = black] (-2.85cm, 0.525cm) -- (3.4cm, 0.525cm);
    	\draw [draw = black] (-2.85cm, 0.525cm) -- (-2.85cm, 2.85cm);
    	\draw [draw = black] (3.4cm, 0.525cm) -- (3.4cm, 2.85cm);
	\end{scope}

\end{tikzpicture}
\end{center}
\caption{
\label{fig:spectrum}
Spectrum of the $X (v=0) \leftarrow A (v'=0)$ $Q$-transitions for the two isotopologues Al$^{35}$Cl and Al$^{37}$Cl. The blue dots are the experimental data and the red solid curve is a multi-peak Gaussian fit to extract the peak centers.
   The inverted, solid black line is the {\it ab initio} calculation, which reproduces the experimental isotope shift within $\Delta \approx 4$\,\%, see text for details.
   \textbf{Inset:} Measured isotope shifts of the Al$^{35}$Cl and Al$^{37}$Cl peaks marked in the main plot (a,\ldots,g). The average value (dashed line) is $6124 \pm 8$\,MHz.
}
\end{figure}
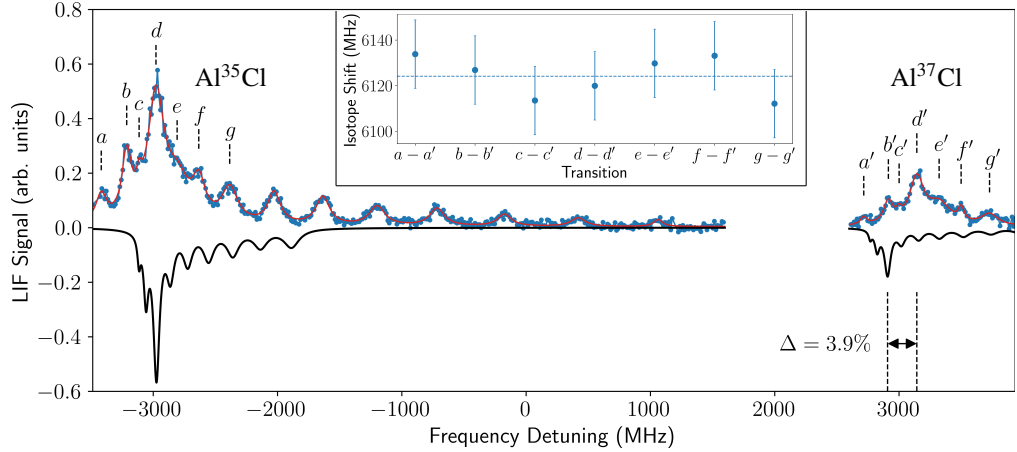

Using the method described in \rsec{sec:technique}, we carried out detailed spectroscopy on the isotope shift of the $X^1\Sigma (v = 0) \leftarrow A^1\Pi (v' = 0)$ $Q$-transitions at 261.5\,nm in AlCl.
\rfig{fig:spectrum} shows the laser-induced fluorescence spectrum of both isotopologues, Al$^{35}$Cl and Al$^{37}$Cl, as a function of the laser detuning. The blue dots are the measured data and each point corresponds to 30 averages, and the red solid curve corresponds to a Gaussian multi-peak fit to identify the center of each line. 
This data was taken using the calibrated wavelength meter, as described above. The difference in the amplitudes of two isotope manifolds comes from the natural abundances of 75.8\% for $^{35}$Cl and 24.2\% for $^{37}$Cl.
The individual peaks of each manifold are a combination of the $Q(1), Q(2), \ldots$-transitions and the hyperfine structure splittings of the involved states that originates from the nuclear spins of Al ($I_1=5/2$) and $^{35/37}$Cl ($I_2=3/2$) \cite{Daniel2021, Daniel2023}.
The short lifetime of $6$\,ns of the $A^1\Pi$ electronic state results in a broad natural linewidth ($\approx 27$\,MHz) that does not resolve the chlorine hyperfine splitting which spreads over $\approx 11.5$\,MHz \cite{Hensel1993}.
The similarity of the rotational constants of the $X^1\Sigma$- and $A^1\Pi$-states \cite{Daniel2021} results in an overlap of the first few $Q$ transitions.
Therefore, we chose to compare the frequency of the fitted combined peaks, labeled $a,b, \dots,g$ and $a',b', \dots, g'$, to extract an average isotope shift.
The result of this comparison is shown in the inset of \rfig{fig:spectrum} for all seven peaks with an average of $6124 \pm 8$\,MHz.
\change{
This measurement is consistent with the isotope shift extracted from the lower resolution absorption spectroscopy in our previous work, which yielded $\approx 6128$\,MHz \cite{Daniel2021} and is the first report of the electronic isotope shift in AlCl of the $X (v=0) \leftarrow A (v'=0)$ $Q$-transition.
To the best of our knowledge, the only available other detailed work on isotope shifts in AlCl is Nakadi et al.~\cite{Nakadi2015}. 
Due to their instrument's resolution of $\approx 1.5$\,pm, they were not able to observe this transition and reported measurements only for higher vibrational states ($v \ge 1$).
The characterization of the $X (v=0) \leftarrow A (v'=0)$ transition, however, is important for future atomic physics experiments with AlCl since it is part of the laser cooling scheme of AlCl.
}

The theoretical calculations for AlCl are discussed in detail in our prior work\cite{Daniel2021}, so only a brief summary will be given here.
Accurate {\it ab initio} electronic structure calculations were performed using MOLPRO \cite{Werner2015} to compute the potential energy curves for the ground singlet $X\,^1\Sigma^+$, excited singlet $A\,^1\Pi_1$ and triplet $a\,^3\Pi$ states.  These potential energy curves were then used to perform a numerically exact solution of the one-dimensional diatomic rovibrational Schr\"odinger equation for the rovibrational wave functions and energies.
The relevant transition frequencies for the $R$, $Q$ and $P$ branches for the $A\, ^1\Pi \leftarrow X\, ^1\Sigma^+$ transition were then calculated. Overall, excellent agreement was obtained between the theoretical and experimental frequencies \cite{Daniel2021}.
In the present work, the rotational structure within the $Q$ branch is computed for both Al$^{35}$Cl and Al$^{37}$Cl for the $X\, ^1\Sigma^+  \leftarrow A\, ^1\Pi$ transition.
The isotope shift between the two $Q$ branches is also calculated enabling a direct comparison with the experimental
spectra and isotope shift presented in \rfig{fig:spectrum}.
\rtab{table1} lists the experimental frequencies for the various features (fitted peaks) labeled by $a$ to $g$
for Al$^{35}$Cl and by $a'$ to $g'$ for Al$^{37}$Cl in \rfig{fig:spectrum}.
The isotope shift for each feature is also tabulated.
The corresponding theoretical frequencies for each $Q$ transition (i.e., $Q(1)$, $Q(2)$, $\ldots$) are also listed for
each isotopologue as well as the associated isotope shift.

\begin{table}
 \caption{
   Q-branch transition frequencies (MHz) are tabulated for X($v=0$) $\leftarrow$ A($v'=0$) in Al$^{35}$Cl and Al$^{37}$Cl. The isotope shift is computed for each transition. The theoretical {\it ab initio} results are also tabulated. The experimental and theoretical average isotope shifts differ by $4\%$ and are  $6124 \pm 8$ and $5881$ MHz, respectively. All frequencies are relative to the center frequency of $1146.33415$ THz.
   }
\begin{tabular}{cc|ccc|ccc}
\hline
\hline
   &&
   \multicolumn{3}{c|}{\small Experimental} 
   &
   \multicolumn{3}{c}{\small Theoretical}\\
\hline
\small Label & 
   \small Q(i) &
  \small Al$^{35}$Cl &
  \small Al$^{37}$Cl &
  \small Isotope Shift &
  \small Al$^{35}$Cl &
  \small Al$^{37}$Cl &
  \small Isotope Shift  \\
\hline
\hline
a,a' & 1  & -3417   & 2717 & 6134  & -3113    & 2769  &  5882 \\
b,b' & 2  & -3212   & 2915 & 6127  & -3057    & 2825  &  5882 \\
c,c' &   & -3112   & 3001  & 6114  &          &       &       \\
d,d' & 3  & -2976   & 3144 & 6120  & -2973    & 2909  &  5882 \\
e,e' &   & -2807   & 3323  & 6130  &          &       &       \\
f,f' & 4  & -2632   & 3501 & 6133  & -2861    & 3020  &  5881 \\
g,g' & 5  & -2383   & 3729 & 6112  & -2721    & 3159  &  5880 \\
\hline
\hline
\end{tabular}
\label{table1}
\end{table}

\change{
The differences between the experimental and theoretical Q line centers are quite reasonable considering that no adjustable parameters were optimized to fit the theoretically computed rotational spectra to the experimentally measured one \cite{Daniel2023}. The average theoretical isotope
shift is $5881$\,MHz which is within $4$\% of the experimental value. This level of agreement is considered excellent for {\it ab initio} based theory with no adjustable parameters \cite{Daniel2023}.
}
We note that the $c,c'$ and  $e,e'$ features observed in the experimental spectra are most likely associated with hyperfine structure within the overlapping $Q$ transitions and is not included in the current level of theory.
The theoretical frequencies listed in \rtab{table1} are used to simulate the experimental spectra by adding up the appropriately weighted contributions from each $Q$ transition with a Lorentzian line-width function \cite{Bernath2005} centered at each transition frequency.
The line widths were chosen to increase linearly with the rotational quantum number $J$ via  $15\,(J+1)$ MHz to qualitatively reproduce the experimentally observed widths of the $Q$ peaks in \rfig{fig:spectrum}.
The weights of each $Q$ transition were chosen to match the relative intensities of the experimental spectra for Al$^{35}$Cl in \rfig{fig:spectrum}.
The weights for the Al$^{37}$Cl $Q$ transitions were set by scaling the Al$^{35}$Cl weights by the relative isotope abundance ratio: $0.24/0.76 = 0.32$.

Overall, the theoretically simulated spectra presented in \rfig{fig:spectrum} reproduces the experimental one quite well confirming the assignments of the primary overlapping $Q$ branch features.
The theoretical analysis also confirms that the experimentally measured isotope shift is due to an increase in the relative difference between the vibrational zero point energies (ZPE)
for the $X$ and $A$-states upon isotopic substitution.
That is, the vibrational ZPEs for the $X$ and $A$ states of Al$^{35}$Cl are $240.047\,{\rm cm}^{-1}$ and $224.751\,{\rm cm}^{-1}$, respectively.
Whereas for Al$^{37}$Cl the ZPEs for the $X$ and $A$ states are $237.213\,{\rm cm}^{-1}$ and $222.108\,{\rm cm}^{-1}$.
The corresponding isotope shifts in the vibrational ZPEs for the $X$ and $A$ states are therefore $-2.834\,{\rm cm}^{-1}$ and $-2.643\,{\rm cm}^{-1}$.
Thus, we see that the ZPE for the $X$ state shifts lower in energy than that for the $A$ state and the relative difference increases by $0.191\,{\rm cm}^{-1}$ (or $5726$\,MHz).
The increased relative difference in vibrational ZPE accounts for $97\%$ of the observed isotope shift ($5726/5881 = 0.97$).
The remaining $3\%$ is due to shifts in the rotational energies due to changes in the rotational constants upon isotopic substitution.
We note that the difference in the vibrational ZPE shifts between the $A$ and $X$ states is due to the difference in the potential energy curves (e.g., force constants).
Also, the electronic transition energy $T_e$ is not affected by isotopic substitution and is the same for both isotopologues.


\subsection{Characterization of Long-Term Behaviour and Further Improvements}
\label{sec:improvements}

\begin{figure}[t]
\begin{center}
\includegraphics[width=0.9\linewidth]{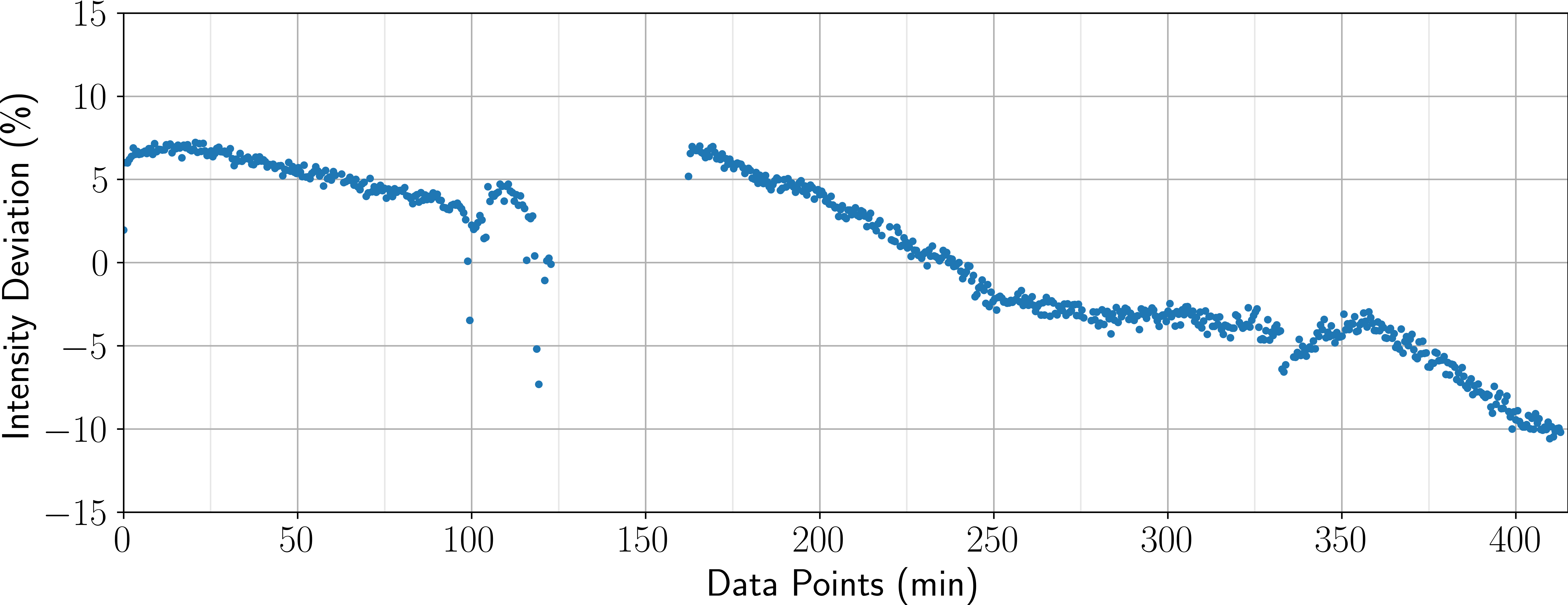}
\end{center}
\caption{
\label{fig:intensity_drift}
Drift of the average UV output intensity over the course of $\approx 6.5$ hours. At $\approx 120$\,min, the experiment was stopped and restarted at $\approx 160$\,min.
}
\end{figure}

\change{
Using our technique, the measurements in \rfig{fig:spectrum} were conducted over a two-day period.}
The primary reason for the long duration of the experimental run is the low repetition rate of $\approx 1$\,Hz and the high number of averages per data point. The experiment is run at such a slow rate to keep the heat load from the ablation laser on the cryogenic cell at a minimum, as is typically done in CBGB experiments. High-number averaging is necessary to compensate for variations in the molecular yield when the ablation laser rasters over the target.

To characterize the long-term behaviour of our method, we monitored the output intensity over a long period of time, as shown in \rfig{fig:intensity_drift}. The average intensity variations between frequency points in the scans were mostly below $10\%$ for at least 6\,hours without any cavity alignment optimization. 
\change{
We attribute these fluctuations primarily to set point drifts in the SHG cavity stabilization circuitry and the temperature sensitivity of polarizing optical components.
Overall, the drifts are comparable to what has been observed in other systems \cite{Hannig2018, Tricot2018}.
However, while these fluctuations introduce no systematic frequency offsets in the experiment, an improved power stability could be achieved by incorporating an acousto-optic modulator at the output of the cavity for intensity stabilization \cite{Tricot2018}. 
}

We note that the SHG cavity can drift in and out of resonance outside the experimental sequence window when the piezo is held at a constant voltage.
This effect leads to unwanted threshold trigger events that switch on the cavity feedback loop. A simple way to avoid this issue is to only arm the threshold trigger at the beginning of the applied voltage ramp. Moreover, if it is desirable to avoid that the cavity drifts into resonance at any time, an easy mitigation technique would be to monitor the output intensity outside the sequence window and use another threshold trigger in combination with an inverted PI control to deliberately push the cavity out of resonance.
For the data presented in this work, we didn't restrict the trigger window and relied on occasional manual adjustments of the offset voltage of the ramp
\change{which was sufficient to run the experiment.}

To increase the UV output power, e.g.~for laser cooling purposes, in our home built system, we also explored using a CLBO crystal (\company{Conex Optics Inc.}) for the SHG conversion. This choice of crystal is motivated by the fact that CLBO has an overall higher conversion efficiency than BBO at our fundamental wavelength of 523\,nm, which is mainly attributed to the reduced walk-off angle of CLBO \cite{McCarron2021,Zhou1996}.
Overall, the cavity UV output increased by an order of magnitude using the CLBO.
However, when we used the cavity in continuous mode, i.e.~before the development of the pulsed technique, we observed a severe degradation of the crystal performance over the course of minutes.
The cavity output power could be brought back to a maximum by translating the crystal, which in turn moves the focus and the entry and exit points of the fundamental cavity mode to a different spot. Translating the focus back to the original point resulted in the low output power again.
\change{
This behaviour was observed both with the BBO and CLBO crystals.
}
In addition, in the case of CLBO, we observed fractures in the crystal that developed over the course of several days of operation, as shown in \rfig{fig:crystal}.

\change{
We attribute these damages to operating the non-linear crystals at room temperature and to their high sensitivity to the presence of humidity in the air near the crystal.
For both crystal types, the negative effects of elevated humidity have been studied extensively.
In the case of CLBO, it has been observed that in particular humidity levels higher than 40\% are detrimental to the crystal's performance \cite{Pan2002,Yuan2006}. While at lower levels, water still gets incorporated in the crystal structure but does not degrade the crystal performance, at levels above 40\% water reacts with the CLBO itself to form polycrystals at the surface that have a different crystal phase \cite{Seryotkin2013}. Moreover,  exposing a hydrated crystal with deep-UV light at 266\,nm leads to the formation of OH groups, which in turn leads to damaging photo-assisted chemical reactions on the crystal surface \cite{Morimoto2001}.
Finally, studies have shown that water molecules preferably penetrate the facets parallel to the CLBO's c-axis, leading to cracks due to hydrolysis within days when exposed to humidity levels higher than 40\% \cite{Pan2002,Yuan2006}.
For BBO, a two-photon absorption process in the deep UV is suggested to be responsible for creating color centers. These defects lead to localized heating due to absorption at longer wavelengths, which reduces the conversion efficiency due to localized changes of the phase-matching condition \cite{Dubietis2000,Isaenko2001,Kondratyuk2002,Divall2005,Kurdi2005,Takahashi2011,Bhandari2012,Kumar2015}.

Consequently, it is recommended to keep such non-linear crystals in a dry environment and use high-quality crystals since internal defects are known to form cracks more easily, at least for CLBO \cite{Pan2002}.
More importantly, the hydration and degradation of the non-linear crystals can be drastically reduced when keeping the crystal at high temperatures. For CLBO, operating temperatures of 150$^\circ$\,C have shown to mitigate the degradation effects and to allow for continuous operation of a high-power SHG cavity \cite{Yap1998,McCarron2021}. Equivalently, for BBO at operating temperature of 200$^\circ$\,C, a 3.5-fold reduction of the two-photon absorption coefficient has been observed, allowing for continuous operation of a SHG cavity \cite{Kumar2015}.

While our enclosure of our SHG cavity provides a clean and relatively dry environment during the experiment, any manual alignment of the cavity happens when the enclosure is open which exposes the crystals to the ambient laboratory air with humidity levels often much higher than 40\%.
This severely limits the ability to improve the cavity alignment to improve the UV output power and we assume that this external water exposure is the most likely cause of the cracked CLBO that we observed.
In addition, we observed that the operation of the SHG cavity in continuous mode seemingly accelerated the degradation effect of the crystal. 
On the other hand, our developed pulsed technique allowed us to operate the homebuilt SHG cavity for a prolonged time even at room temperature without any notable crystal damage.
It is likely that a side effect of reducing the UV-duty cycle is that any detrimental photo-assisted effects and damages in the crystal are reduced.
}

\change{
In the future, we plan to add a temperature control to maintain the non-linear crystal at high temperatures above 130\,$^\circ$ C at all times to mitigate these challenges and develop a laser system at 260\,nm with Watt-level output powers.
Here, reducing the duty cycle of the UV light will be paramount to protect any involved optical elements after the SHG cavity. However, while we expect that the elevated temperature of the crystal removes the hydration and corresponding degradation issues, there are several aspects that need to be considered when employing the pulsed technique in a high power setting.

Any technique that runs an experimental setup with a reduced duty cycle exhibits pulsed heat loads, resulting in a non-equilibrated state of the setup.
In this case of a SHG cavity, the absorption of the light in the crystal leads to what is known as self-heating, which has been observed with both BBO and CLBO crystals, especially in high-power UV systems \cite{Cooper2018,Zhang2021c,Kaneda2016}. 
As a consequence, the phase matching condition is slightly altered due to different net heat loads between the configuration when the user aligns the cavity setup for optimal output power and when the setup is used for an experimental run. This change leads to a reduction of the conversion efficiency and can be mitigated in different ways.
In one approach, users align and optimize the SHG cavities in a configuration that mimics the experimental run in terms of, for instance, repetition rates and used laser powers as close as possible to operate under the same heat loads.
In a second approach, users intentionally lower the temperature of their crystals, slightly deviating from the optimal phase matching condition, as has been shown in previous works \cite{Zhang2021c,Kaneda2016,Burkley2019}.
Yet another way, which we plan to employ, is to maintain the crystal at an elevated temperature such that the relative temperature increase due to self-heating is small compared to the crystal temperature. The latter approach is best applied to crystals that use angle, instead of temperature, crystal phase-matching, which is the case for BBO and CLBO around 500\,nm.

Finally, one expects that the thermal shock of the laser absorption deforms the lattice of the nonlinear crystal, eventually leading to a decrease of the conversion efficiency. This effect, however, has been studied in detail in CLBO and it has been shown that when operating at temperature above 130\,deg C long-term operation is possible since the thermal stress is minimized due to activated phonon vibration \cite{Yap1998}.
}


\section{Conclusion}

\change{
In this work, we have developed and tested a technique to reduce the duty cycle of a deep-UV SHG laser system.
Our approach is ideal for any pulsed experiments but can also be applied in CW experiments to bridge, often unavoidable, dead times and reduce the UV operation time as much as possible.
}
The frequency accuracy of the laser system should not be affected when using this method since the ramping of the piezo-actuated mirror only affects the output intensity and the frequency stability remains with the fundamental laser.
The long-term stability of our setup is sufficient for typical time scales of atomic and molecular experiments.
We applied this technique to measure the isotope shift of the electronic $X^1\Sigma$-$A^1\Pi$ transition in AlCl and found excellent agreement with {\it ab initio} calculations.

\change{
Our technique is particularly well-suited in high-power deep-UV applications to prolong the lifetime of any optical component, such as mirrors, optical fibers or waveplates, used to guide and manipulate UV light in an experimental setup.
For example, laser cooling schemes for molecules, such as AlCl, require high laser powers due to its high saturation intensity of 232\,mW/cm$^2$ \cite{Daniel2023}.
Typical laser cooling setups often use, e.g., acousto-optic modulators in a double pass configuration to implement frequency shifts and chirps of the laser light while maintaining the beam alignment \cite{Chang2008, Donley2005}.
These setups are particularly challenging in the UV since they require focusing the beam on a retroreflector \cite{Hemmerling2011}.
With a 300\,mm focal length lens with a 500$\mu$m beam diameter for a 2\,W laser beam yields a linear power density of 400\,W/cm, which reaches the specified damage threshold of typical commercially available optics.
We expect that our approach will be beneficial to minimize UV-induced damage in such systems.
}


\begin{backmatter}
\bmsection{Funding}
L.~L.~and B.~H.~acknowledge funding from the National Science Foundation under Grant No.~2145147.
This material is based upon work supported by the Air Force Office of Scientific Research under award number FA9550-21-1-0263. B.~K.~K.~acknowledges that part of this work was done under the auspices of the U.S.~Department of Energy under Project No. 20240256ER of the Laboratory Directed Research and Development Program at Los Alamos National Laboratory. Los Alamos National Laboratory is operated by Triad National Security, LLC, for the National Nuclear Security Administration of the U.S.~Department of Energy (Contract No.~89233218CNA000001).

\bmsection{Acknowledgments}
We would like to thank Grady Kestler and Julio T.~Barreiro for helpful discussions on the FPGA implementation and the PyRPL code.
We would also like to thank Daniel J.~McCarron for useful discussions and feedback on this manuscript.

\bmsection{Disclosures}
The authors declare no conflicts of interest.

\bmsection{Data availability}
Data underlying the results presented in this paper are available upon request.

\end{backmatter}


\label{sec:refs}

\bibliography{library}

\begin{thebibliography}{100}
\newcommand{\enquote}[1]{``#1''}

\bibitem{Duarte2016}
F.~Duarte, \emph{Tunable Laser Applications}, ISSN (CRC Press, 2016).

\bibitem{Safronova2018}
M.~Safronova, D.~Budker, D.~DeMille, \emph{et~al.}, \enquote{Search for new
  physics with atoms and molecules,} {\protect\JournalTitle{Reviews of Modern
  Physics}} \textbf{90}, 025008 (2018).

\bibitem{Metcalf1999}
M.~H. J and van~der Straten~P, \emph{Laser Cooling and Trapping} (New York:
  Springer, 1999).

\bibitem{McCarron2018a}
D.~McCarron, \enquote{Laser cooling and trapping molecules,}
  {\protect\JournalTitle{Journal of Physics B: Atomic, Molecular and Optical
  Physics}} \textbf{51}, 212001 (2018).

\bibitem{Krems2008}
R.~V. Krems, \enquote{Cold controlled chemistry,}
  {\protect\JournalTitle{Physical Chemistry Chemical Physics}} \textbf{10},
  4079 (2008).

\bibitem{Carr2009}
L.~D. Carr, D.~DeMille, R.~V. Krems, and J.~Ye, \enquote{Cold and ultracold
  molecules: science, technology and applications,} {\protect\JournalTitle{New
  Journal of Physics}} \textbf{11}, 055049 (2009).

\bibitem{Krems2019}
R.~V. Krems, B.~Friedrich, and W.~C. Stwalley, \emph{Cold molecules: theory,
  experiment, applications.} (CRC PRESS, 2019).

\bibitem{Brickman2007}
K.-A. Brickman, M.-S. Chang, M.~Acton, \emph{et~al.}, \enquote{Magneto-optical
  trapping of cadmium,} {\protect\JournalTitle{Physical Review A}} \textbf{76},
  043411 (2007).

\bibitem{Kaneda2016}
Y.~Kaneda, J.~M. Yarborough, Y.~Merzlyak, \emph{et~al.},
  \enquote{Continuous-wave, single-frequency 229 nm laser source for laser
  cooling of cadmium atoms,} {\protect\JournalTitle{Optics Letters}}
  \textbf{41}, 705 (2016).

\bibitem{Tinsley2021}
J.~N. Tinsley, S.~Bandarupally, J.-P. Penttinen, \emph{et~al.},
  \enquote{Watt-level blue light for precision spectroscopy, laser cooling and
  trapping of strontium and cadmium atoms,} {\protect\JournalTitle{Optics
  Express}} \textbf{29}, 25462--25476 (2021).

\bibitem{Villwock2011}
P.~Villwock, S.~Siol, and T.~Walther, \enquote{Magneto-optical trapping of
  neutral mercury,} {\protect\JournalTitle{The European Physical Journal D}}
  \textbf{65}, 251--255 (2011).

\bibitem{Zhang2021c}
Y.~Zhang, Q.~Liu, X.~Fu, \emph{et~al.}, \enquote{A stable deep-ultraviolet
  laser for laser cooling of mercury atoms,} {\protect\JournalTitle{Optics \&
  Laser Technology}} \textbf{139}, 106956 (2021).

\bibitem{Zhang2019}
Z.~Zhang, M.~Kushimoto, T.~Sakai, \emph{et~al.}, \enquote{A 271.8 nm
  deep-ultraviolet laser diode for room temperature operation,}
  {\protect\JournalTitle{Applied Physics Express}} \textbf{12}, 124003 (2019).

\bibitem{Amano2020}
H.~Amano, R.~Collazo, C.~D. Santi, \emph{et~al.}, \enquote{The 2020 uv emitter
  roadmap,} {\protect\JournalTitle{Journal of Physics D: Applied Physics}}
  \textbf{53}, 503001 (2020).

\bibitem{Parthey2011}
C.~G. Parthey, A.~Matveev, J.~Alnis, \emph{et~al.}, \enquote{Improved
  {Measurement} of the {Hydrogen} 1s--2s {Transition} {Frequency},}
  {\protect\JournalTitle{Physical Review Letters}} \textbf{107}, 203001 (2011).

\bibitem{Ahmadi2018}
M.~Ahmadi, B.~X.~R. Alves, C.~J. Baker, \emph{et~al.},
  \enquote{Characterization of the {1S}–{2S} transition in antihydrogen,}
  {\protect\JournalTitle{Nature}} \textbf{557}, 71--75 (2018).

\bibitem{Fleurbaey2018}
H.~Fleurbaey, S.~Galtier, S.~Thomas, \emph{et~al.}, \enquote{New {Measurement}
  of the 1s-3s {Transition} {Frequency} of {Hydrogen}: {Contribution} to the
  {Proton} {Charge} {Radius} {Puzzle},} {\protect\JournalTitle{Physical Review
  Letters}} \textbf{120}, 183001 (2018).

\bibitem{Grinin2020}
A.~Grinin, A.~Matveev, D.~C. Yost, \emph{et~al.}, \enquote{Two-photon frequency
  comb spectroscopy of atomic hydrogen,} {\protect\JournalTitle{Science}}
  \textbf{370}, 1061--1066 (2020).

\bibitem{Crivelli2018}
P.~Crivelli, \enquote{The {Mu}-{MASS} (muonium laser spectroscopy) experiment,}
  {\protect\JournalTitle{Hyperfine Interactions}} \textbf{239}, 49 (2018).

\bibitem{Altiere2018}
E.~Altiere, E.~R. Miller, T.~Hayamizu, \emph{et~al.}, \enquote{High-resolution
  two-photon spectroscopy of a $5p^4 6p\leftarrow 5p^6$ transition of xenon,}
  {\protect\JournalTitle{Physical Review A}} \textbf{97}, 012507 (2018).

\bibitem{Rosa2004}
M.~D.~D. Rosa, \enquote{Laser-cooling molecules,} {\protect\JournalTitle{The
  European Physical Journal D}} \textbf{31}, 395--402 (2004).

\bibitem{Tarbutt2019}
M.~R. Tarbutt, \enquote{Laser cooling of molecules,}
  {\protect\JournalTitle{Contemporary Physics}} \textbf{59}, 356--376 (2018).

\bibitem{Chae2023}
E.~Chae, \enquote{Laser cooling of molecules,} {\protect\JournalTitle{Journal
  of the Korean Physical Society}} \textbf{82}, 851--863 (2023).

\bibitem{Daniel2021}
J.~R. Daniel, C.~Wang, K.~Rodriguez, \emph{et~al.}, \enquote{Spectroscopy on
  the x$^1\sigma^+$ -- a$^1\pi$ transition of buffer-gas cooled alcl,}
  {\protect\JournalTitle{Physical Review A}} \textbf{104}, 012801 (2021).

\bibitem{Daniel2023}
J.~R. Daniel, J.~C. Shaw, C.~Wang, \emph{et~al.}, \enquote{Hyperfine structure
  of the a$^1\pi$ state of alcl and its relevance to laser cooling and
  trapping,} {\protect\JournalTitle{Physical Review A}} \textbf{108}, 062821
  (2023).

\bibitem{Hofsass2021}
S.~Hofsäss, M.~Doppelbauer, S.~C. Wright, \emph{et~al.}, \enquote{Optical
  cycling of {AlF} molecules,} {\protect\JournalTitle{New Journal of Physics}}
  \textbf{23}, 075001 (2021).

\bibitem{Mes2003b}
J.~Mes, E.~J. van Duijn, R.~Zinkstok, \emph{et~al.}, \enquote{Third-harmonic
  generation of a continuous-wave ti:sapphire laser in external resonant
  cavities,} {\protect\JournalTitle{Applied Physics Letters}} \textbf{82},
  4423--4425 (2003).

\bibitem{McCarron2021}
J.~C. Shaw, S.~Hannig, and D.~J. McCarron, \enquote{Stable 2 w continuous-wave
  261.5 nm laser for cooling and trapping aluminum monochloride,}
  {\protect\JournalTitle{Optics Express}} \textbf{29}, 37140--37149 (2021).

\bibitem{Zhadnov2023}
N.~Zhadnov, A.~Golovizin, I.~Cortinovis, \emph{et~al.}, \enquote{Pulsed {CW}
  laser for long-term spectroscopic measurements at high power in deep-{UV},}
  {\protect\JournalTitle{Optics Express}} \textbf{31}, 28470--28479 (2023).

\bibitem{Burkley2019}
Z.~Burkley, A.~D. Brandt, C.~Rasor, \emph{et~al.}, \enquote{Highly coherent,
  watt-level deep-uv radiation via a frequency-quadrupled yb-fiber laser
  system,} {\protect\JournalTitle{Applied Optics}} \textbf{58}, 1657--1661
  (2019).

\bibitem{Hu2013}
J.~Hu, L.~Zhang, H.~Liu, \emph{et~al.}, \enquote{High power room temperature
  1014.8 nm {Yb} fiber amplifier and frequency quadrupling to 253.7 nm for
  laser cooling of mercury atoms,} {\protect\JournalTitle{Optics Express}}
  \textbf{21}, 30958--30963 (2013).

\bibitem{Sayama1997}
S.~Sayama and M.~Ohtsu, \enquote{Tunable uv cw generation by frequency tripling
  of a ti:sapphire laser,} {\protect\JournalTitle{Optics Communications}}
  \textbf{137}, 295--298 (1997).

\bibitem{Cooper2018}
S.~F. Cooper, Z.~Burkley, A.~D. Brandt, \emph{et~al.}, \enquote{Cavity-enhanced
  deep ultraviolet laser for two-photon cooling of atomic hydrogen,}
  {\protect\JournalTitle{Optics Letters}} \textbf{43}, 1375 (2018).

\bibitem{Imai2003}
S.~Imai, H.~Inoue, T.~Nomura, and T.~Tojo, \enquote{Cw 198.5-nm light
  generation in clbo,} {\protect\JournalTitle{Advanced Solid-State Photonics}}
  p. 380 (2003).

\bibitem{Asakawa2004}
Y.~Asakawa, J.~Sakuma, H.~Sekita, and M.~Obara, \enquote{High-power cw duv
  coherent light source around 200 nm,} {\protect\JournalTitle{Advanced
  Solid-State Photonics (TOPS)}} p. PDP12 (2004).

\bibitem{Huang2008}
S.~Huang, T.~Ando, Y.~Orii, and T.~Sumiyoshi, \enquote{A cw 266 nm coherent
  light source pumped by the shg beam of the single frequency fiber amplifier
  radiation,} {\protect\JournalTitle{Advanced Solid-State Photonics}} p. MC25
  (2008).

\bibitem{Oka2004}
M.~Oka, L.~Liu, W.~Wiechmann, \emph{et~al.}, \enquote{1 w continuous-wave 266
  nm radiation from an all solid-state frequency quadrupled nd:yag laser,}
  {\protect\JournalTitle{Advanced Solid State Lasers}} p. US1 (2004).

\bibitem{Masuda2001}
H.~Masuda, K.~Kimura, N.~Eguchi, \emph{et~al.}, \enquote{All-solid-state,
  continuous-wave, 195 nm light generation in {β-BaB2O4},}
  {\protect\JournalTitle{Advanced Solid-State Lasers}} p. WA6 (2001).

\bibitem{Bauer2009}
M.~Bauer, M.~Bischoff, S.~Jukresch, \emph{et~al.}, \enquote{Exterior surface
  damage of calcium fluoride outcoupling mirrors for duv lasers,}
  {\protect\JournalTitle{Optics Express}} \textbf{17}, 8253 (2009).

\bibitem{Bauer2009a}
M.~Bauer, M.~Bischoff, T.~H\"{u}lsenbusch, \emph{et~al.}, \enquote{Onset of the
  optical damage in caf optics caused by deep-uv lasers,}
  {\protect\JournalTitle{Optics Letters}} \textbf{34}, 3815 (2009).

\bibitem{Schenker1994}
R.~Schenker, P.~Schermerhorn, and W.~G. Oldham, \enquote{Deep‐ultraviolet
  damage to fused silica,} {\protect\JournalTitle{Journal of Vacuum Science \&
  Technology B: Microelectronics and Nanometer Structures Processing,
  Measurement, and Phenomena}} \textbf{12}, 3275--3279 (1994).

\bibitem{Schenker1995}
R.~E. Schenker, L.~Eichner, H.~Vaidya, \emph{et~al.}, \enquote{Ultraviolet
  damage properties of various fused silica materials,}
  {\protect\JournalTitle{Laser-Induced Damage in Optical Materials: 1994}}
  \textbf{2428}, 458--468 (1995).

\bibitem{Negres2010}
R.~A. Negres, M.~A. Norton, D.~A. Cross, and C.~W. Carr, \enquote{Growth
  behavior of laser-induced damage on fused silica optics under {UV}, ns laser
  irradiation,} {\protect\JournalTitle{Optics Express}} \textbf{18},
  19966--19976 (2010).

\bibitem{Kunz2000}
R.~R. Kunz, V.~Liberman, and D.~K. Downs, \enquote{Experimentation and modeling
  of organic photocontamination on lithographic optics,}
  {\protect\JournalTitle{Journal of Vacuum Science \& Technology B:
  Microelectronics and Nanometer Structures Processing, Measurement, and
  Phenomena}} \textbf{18}, 1306--1313 (2000).

\bibitem{Heinbuch2008}
S.~Heinbuch, F.~Dong, J.~J. Rocca, and E.~R. Bernstein, \enquote{Gas-phase
  study of the reactivity of optical coating materials with hydrocarbons by use
  of a desktop-size extreme-ultraviolet laser,} {\protect\JournalTitle{Journal
  of the Optical Society of America B}} \textbf{25}, B85--B91 (2008).

\bibitem{Gangloff2015}
D.~Gangloff, M.~Shi, T.~Wu, \emph{et~al.}, \enquote{Preventing and reversing
  vacuum-induced optical losses in high-finesse tantalum ({V}) oxide mirror
  coatings,} {\protect\JournalTitle{Optics Express}} \textbf{23}, 18014--18028
  (2015).

\bibitem{Burkley2021}
Z.~Burkley, L.~d.~S. Borges, B.~Ohayon, \emph{et~al.}, \enquote{Stable high
  power deep-uv enhancement cavity in ultra-high vacuum with fluoride
  coatings,} {\protect\JournalTitle{Optics Express}} \textbf{29}, 27450--27459
  (2021).

\bibitem{Watanabe1991}
M.~Watanabe, K.~Hayasaka, H.~Imajo, \emph{et~al.}, \enquote{Generation of
  continuous-wave coherent radiation tunable down to 190.8nm in β-{BaB2O4},}
  {\protect\JournalTitle{Applied Physics B}} \textbf{53}, 11--13 (1991).

\bibitem{Takahashi2010}
M.~Takahashi, A.~Osada, A.~Dergachev, \emph{et~al.}, \enquote{Effects of
  {Pulse} {Rate} and {Temperature} on {Nonlinear} {Absorption} of {Pulsed}
  262-nm {Laser} {Light} in β-{BaB2O4},} {\protect\JournalTitle{Japanese
  Journal of Applied Physics}} \textbf{49}, 080211 (2010).

\bibitem{Turcicova2022}
H.~Turcicova, O.~Novak, J.~Muzik, \emph{et~al.}, \enquote{Laser induced damage
  threshold ({LIDT}) of β-barium borate ({BBO}) and cesium lithium borate
  ({CLBO}) – {Overview},} {\protect\JournalTitle{Optics \& Laser Technology}}
  \textbf{149}, 107876 (2022).

\bibitem{Nishioka2005}
M.~Nishioka, A.~Kanoh, M.~Yoshimura, \emph{et~al.}, \enquote{Improvement in
  {UV} {Optical} {Properties} of {CsLiB6O10} by {Reducing} {Water} {Molecules}
  in the {Crystal},} {\protect\JournalTitle{Japanese Journal of Applied
  Physics}} \textbf{44}, L699 (2005).

\bibitem{Kawamura2009}
T.~Kawamura, M.~Yoshimura, Y.~Honda, \emph{et~al.}, \enquote{Effect of water
  impurity in cslib6i10 crystals on bulk laser-induced damage threshold and
  transmittance in the ultraviolet region,} {\protect\JournalTitle{Applied
  Optics}} \textbf{48}, 1658--1662 (2009).

\bibitem{Takachiho2014}
K.~Takachiho, M.~Yoshimura, Y.~Takahashi, \emph{et~al.}, \enquote{Ultraviolet
  laser-induced degradation of {CsLiB}$_{\textrm{6}}${O}$_{\textrm{10}}$ and
  β-{BaB}$_{\textrm{2}}${O}$_{\textrm{4}}$,} {\protect\JournalTitle{Optical
  Materials Express}} \textbf{4}, 559--567 (2014).

\bibitem{Yoshimura2015}
M.~Yoshimura, Y.~Oeki, Y.~Takahashi, \emph{et~al.}, \enquote{Ultraviolet
  laser-induced degradation of cslib6o10,} {\protect\JournalTitle{Advanced
  Solid State Lasers}} p. AM5A.11 (2015).

\bibitem{Colombe2014}
Y.~Colombe, D.~H. Slichter, A.~C. Wilson, \emph{et~al.}, \enquote{Single-mode
  optical fiber for high-power, low-loss uv transmission,}
  {\protect\JournalTitle{Optics Express}} \textbf{22}, 19783--19793 (2014).

\bibitem{Hemmerling2011}
B.~Hemmerling, F.~Gebert, Y.~Wan, \emph{et~al.}, \enquote{A single laser system
  for ground-state cooling of 25mg+,} {\protect\JournalTitle{Applied Physics
  B}} \textbf{104}, 583--590 (2011).

\bibitem{Ni2010}
K.-K.~K. Ni, S.~Ospelkaus, D.~Wang, \emph{et~al.}, \enquote{Dipolar collisions
  of polar molecules in the quantum regime,} {\protect\JournalTitle{Nature}}
  \textbf{464}, 1324--1328 (2010).

\bibitem{Ye2018}
X.~Ye, M.~Guo, M.~L. González-Martínez, \emph{et~al.}, \enquote{Collisions of
  ultracold 23na87rb molecules with controlled chemical reactivities,}
  {\protect\JournalTitle{Science Advances}} \textbf{4}, eaaq0083 (2018).

\bibitem{Ospelkaus2010}
S.~Ospelkaus, K.-K.~K. Ni, D.~Wang, \emph{et~al.}, \enquote{Quantum-state
  controlled chemical reactions of ultracold potassium-rubidium molecules,}
  {\protect\JournalTitle{Science}} \textbf{327}, 853 (2010).

\bibitem{DeMille2002}
D.~DeMille, \enquote{Quantum computation with trapped polar molecules,}
  {\protect\JournalTitle{Physical Review Letters}} \textbf{88}, 067901 (2002).

\bibitem{Yelin2006}
S.~F. Yelin, K.~Kirby, and R.~Côté, \enquote{Schemes for robust quantum
  computation with polar molecules,} {\protect\JournalTitle{Physical Review A}}
  \textbf{74}, 050301(R) (2006).

\bibitem{Yu2019}
P.~Yu, L.~W. Cheuk, I.~Kozyryev, and J.~M. Doyle, \enquote{A scalable quantum
  computing platform using symmetric-top molecules,} {\protect\JournalTitle{New
  Journal of Physics}} \textbf{21}, 093049 (2019).

\bibitem{Micheli2006}
A.~Micheli, G.~K. Brennen, and P.~Zoller, \enquote{A toolbox for lattice-spin
  models with polar molecules,} {\protect\JournalTitle{Nature Physics}}
  \textbf{2}, 341--347 (2006).

\bibitem{Bao2022}
Y.~Bao, S.~S. Yu, L.~Anderegg, \emph{et~al.}, \enquote{Dipolar spin-exchange
  and entanglement between molecules in an optical tweezer array,}
  {\protect\JournalTitle{arXiv:2211.09780}}  (2022). ArXiv:2211.09780 [physics,
  physics:quant-ph] type: article.

\bibitem{Holland2022}
C.~M. Holland, Y.~Lu, and L.~W. Cheuk, \enquote{On-{Demand} {Entanglement} of
  {Molecules} in a {Reconfigurable} {Optical} {Tweezer} {Array},}
  {\protect\JournalTitle{Science}} \textbf{382}, 1143 (2023).

\bibitem{Andreev2018}
V.~Andreev, D.~G. Ang, D.~DeMille, \emph{et~al.}, \enquote{Improved limit on
  the electric dipole moment of the electron,} {\protect\JournalTitle{Nature}}
  \textbf{562}, 355--360 (2018).

\bibitem{Cairncross2017}
W.~B. Cairncross, D.~N. Gresh, M.~Grau, \emph{et~al.}, \enquote{Precision
  measurement of the electron's electric dipole moment using trapped molecular
  ions,} {\protect\JournalTitle{Phys. Rev. Lett.}} \textbf{119}, 153001 (2017).

\bibitem{Kozyryev2017a}
I.~Kozyryev and N.~R. Hutzler, \enquote{Precision measurement of time-reversal
  symmetry violation with laser-cooled polyatomic molecules,}
  {\protect\JournalTitle{Physical Review Letters}} \textbf{119}, 133002 (2017).

\bibitem{Hudson2011}
J.~J. Hudson, D.~M. Kara, I.~J. Smallman, \emph{et~al.}, \enquote{Improved
  measurement of the shape of the electron,} {\protect\JournalTitle{Nature}}
  \textbf{473}, 493--496 (2011).

\bibitem{Kozyryev2021}
I.~Kozyryev, Z.~Lasner, and J.~M. Doyle, \enquote{Enhanced sensitivity to
  ultralight bosonic dark matter in the spectra of the linear radical sroh,}
  {\protect\JournalTitle{Physical Review A}} \textbf{103}, 043313 (2021).

\bibitem{Kondov2019}
S.~S. Kondov, C.-H. Lee, K.~H. Leung, \emph{et~al.}, \enquote{Molecular lattice
  clock with long vibrational coherence,} {\protect\JournalTitle{Nature
  Physics}} \textbf{15}, 1118--1122 (2019).

\bibitem{ACMECollaboration2014}
T.~A.~A. Collaboration, J.~Baron, W.~C. Campbell, \emph{et~al.}, \enquote{Order
  of magnitude smaller limit on the electric dipole moment of the electron.}
  {\protect\JournalTitle{Science}} \textbf{343}, 269--72 (2014).

\bibitem{Fitch2021a}
N.~J. Fitch, J.~Lim, E.~A. Hinds, \emph{et~al.}, \enquote{Methods for measuring
  the electron edm using ultracold ybf molecules,}
  {\protect\JournalTitle{Quantum Sci. Technol.}} \textbf{6}, 014006 (2021).

\bibitem{Yu2021}
P.~Yu and N.~R. Hutzler, \enquote{Probing fundamental symmetries of deformed
  nuclei in symmetric top molecules,} {\protect\JournalTitle{Physical Review
  Letters}} \textbf{126}, 023003 (2021).

\bibitem{Hutzler2020}
N.~R. Hutzler, \enquote{Polyatomic molecules as quantum sensors for fundamental
  physics,} {\protect\JournalTitle{Quantum Science and Technology}} \textbf{5},
  44011 (2021).

\bibitem{ORourke2019}
M.~J. O'Rourke and N.~R. Hutzler, \enquote{Hypermetallic polar molecules for
  precision measurements,} {\protect\JournalTitle{Physical Review A}}
  \textbf{100}, 022502 (2019).

\bibitem{Aggarwal2018}
P.~Aggarwal, H.~L. Bethlem, A.~Borschevsky, \emph{et~al.}, \enquote{Measuring
  the electric dipole moment of the electron in baf,}
  {\protect\JournalTitle{European Physical Journal D}} \textbf{72}, 197 (2018).

\bibitem{Uzan2003}
J.-P. Uzan, \enquote{The fundamental constants and their variation:
  observational and theoretical status,} {\protect\JournalTitle{Reviews of
  Modern Physics}} \textbf{75}, 403 (2003).

\bibitem{DeMille2008}
D.~DeMille, S.~Sainis, J.~Sage, \emph{et~al.}, \enquote{Enhanced sensitivity to
  variation of me/mp in molecular spectra,} {\protect\JournalTitle{Physical
  Review Letters}} \textbf{100}, 043202 (2008).

\bibitem{Chin2009}
C.~Chin, V.~V. Flambaum, and M.~G. Kozlov, \enquote{Ultracold molecules: new
  probes on the variation of fundamental constants,} {\protect\JournalTitle{New
  Journal of Physics}} \textbf{11}, 55048 (2009).

\bibitem{Kajita2009}
M.~Kajita, \enquote{Sensitive measurement of mp/me variance using vibrational
  transition frequencies of cold molecules,} {\protect\JournalTitle{New Journal
  of Physics}} \textbf{11}, 055010 (2009).

\bibitem{Beloy2010}
K.~Beloy, A.~Borschevsky, P.~Schwerdtfeger, and V.~V. Flambaum,
  \enquote{Enhanced sensitivity to the time variation of the fine-structure
  constant and <span class,} {\protect\JournalTitle{Physical Review A}}
  \textbf{82}, 022106 (2010).

\bibitem{Jansen2014}
P.~Jansen, H.~L. Bethlem, and W.~Ubachs, \enquote{Perspective: Tipping the
  scales: Search for drifting constants from molecular spectra,}
  {\protect\JournalTitle{The Journal of Chemical Physics}} \textbf{140}, 010901
  (2014).

\bibitem{Dapra2016}
M.~Daprà, M.~L. Niu, E.~J. Salumbides, \emph{et~al.}, \enquote{Constraint on a
  cosmological variation in the proton-to-electron mass ratio from electroinc
  co absorption,} {\protect\JournalTitle{The Astrophysical Journal}}
  \textbf{826}, 192 (2016).

\bibitem{Kobayashi2019}
J.~Kobayashi, A.~Ogino, and S.~Inouye, \enquote{Measurement of the variation of
  electron-to-proton mass ratio using ultracold molecules produced from
  laser-cooled atoms,} {\protect\JournalTitle{Nature Communications}}
  \textbf{10}, 1--5 (2019).

\bibitem{Chupp2019}
T.~E. Chupp, P.~Fierlinger, M.~J. Ramsey-Musolf, and J.~T. Singh,
  \enquote{Electric dipole moments of atoms, molecules, nuclei, and particles,}
  {\protect\JournalTitle{Reviews of Modern Physics}} \textbf{91}, 015001
  (2019).

\bibitem{Hutzler2012}
N.~R. Hutzler, H.-I.~I. Lu, and J.~M. Doyle, \enquote{The buffer gas beam: An
  intense, cold, and slow source for atoms and molecules,}
  {\protect\JournalTitle{Chemical Reviews}} \textbf{112}, 4803--4827 (2012).

\bibitem{Lewis2021}
T.~N. Lewis, C.~Wang, J.~R. Daniel, \emph{et~al.}, \enquote{Optimizing
  pulsed-laser ablation production of alcl molecules for laser cooling,}
  {\protect\JournalTitle{Physical Chemistry Chemical Physics}} \textbf{23},
  22785 (2021).

\bibitem{Steck2019}
D.~A. Steck, \enquote{Rubidium 85 d line data,}
  {\protect\JournalTitle{http://steck.us/alkalidata}}  (2021).

\bibitem{Kleinman1966}
D.~A. Kleinman, A.~Ashkin, and G.~D. Boyd, \enquote{Second-harmonic generation
  of light by focused laser beams,} {\protect\JournalTitle{Physical Review}}
  \textbf{145}, 338--379 (1966).

\bibitem{Boyd1968}
G.~D. Boyd and D.~A. Kleinman, \enquote{Parametric interaction of focused
  gaussian light beams,} {\protect\JournalTitle{Journal of Applied Physics}}
  \textbf{39}, 3597--3639 (1968).

\bibitem{Hansch1980}
T.~W. Hansch and B.~Couillaud, \enquote{Laser frequency stabilization by
  polarization spectroscopy of a reflecting reference cavity,}
  {\protect\JournalTitle{Optics Communications}} \textbf{35}, 441--444 (2021).

\bibitem{Hannig2018}
S.~Hannig, J.~Mielke, J.~A. Fenske, \emph{et~al.}, \enquote{A highly stable
  monolithic enhancement cavity for second harmonic generation in the
  ultraviolet,} {\protect\JournalTitle{Review of Scientific Instruments}}
  \textbf{89}, 013106 (2018).

\bibitem{Preuschoff2020}
T.~Preuschoff, M.~Schlosser, and G.~Birkl, \enquote{Digital laser frequency and
  intensity stabilization based on the {STEMlab} platform (originally {Red}
  {Pitaya}),} {\protect\JournalTitle{Review of Scientific Instruments}}
  \textbf{91}, 083001 (2020).

\bibitem{Pultinevicius2023}
E.~Pultinevicius, M.~Rockenhäuser, F.~Kogel, \emph{et~al.}, \enquote{A
  scalable scanning transfer cavity laser stabilization scheme based on the
  {Red} {Pitaya} {STEMlab} platform,} {\protect\JournalTitle{Review of
  Scientific Instruments}} \textbf{94}, 103004 (2023).

\bibitem{Lauria2022}
P.~Lauria, W.-T. Kuo, N.~R. Cooper, and J.~T. Barreiro, \enquote{Experimental
  {Realization} of a {Fermionic} {Spin}-{Momentum} {Lattice},}
  {\protect\JournalTitle{Physical Review Letters}} \textbf{128}, 245301 (2022).

\bibitem{Kestler2023}
G.~Kestler, K.~Ton, D.~Filin, \emph{et~al.}, \enquote{State-{Insensitive}
  {Trapping} of {Alkaline}-{Earth} {Atoms} in a {Nanofiber}-{Based} {Optical}
  {Dipole} {Trap},} {\protect\JournalTitle{PRX Quantum}} \textbf{4}, 040308
  (2023).

\bibitem{Neuhaus2024}
L.~Neuhaus, M.~Croquette, R.~Metzdorff, \emph{et~al.}, \enquote{Python red
  pitaya lockbox (pyrpl): An open source software package for digital feedback
  control in quantum optics experiments,} {\protect\JournalTitle{Review of
  Scientific Instruments}} \textbf{95}, 033003 (2024).

\bibitem{Artiq2021}
S.~Bourdeauducq, R.~Jördens, P.~Zotov, \emph{et~al.}, \enquote{Artiq,}
  {\protect\JournalTitle{Zenodo}}  (2021).

\bibitem{Hensel1993}
K.~D. Hensel, C.~Styger, W.~Jäger, \emph{et~al.}, \enquote{Microwave spectra
  of metal chlorides produced using laser ablation,} {\protect\JournalTitle{The
  Journal of Chemical Physics}} \textbf{99}, 3320--3328 (1993).

\bibitem{Nakadi2015}
F.~V. Nakadi, M.~A. M. S.~d. Veiga, M.~Aramendía, \emph{et~al.},
  \enquote{Chlorine isotope determination via the monitoring of the {AlCl}
  molecule by high-resolution continuum source graphite furnace molecular
  absorption spectrometry – a case study,} {\protect\JournalTitle{Journal of
  Analytical Atomic Spectrometry}} \textbf{30}, 1531--1540 (2015).

\bibitem{Werner2015}
H.-J. Werner, P.~J. Knowles, G.~Knizia, \emph{et~al.}, \enquote{{MOLPRO,
  version 2015.1, a package of ab initio programs},} Http://www.molpro.net.

\bibitem{Bernath2005}
P.~F. Bernath, \emph{Spectra of Atoms and Molecules} (Oxford University Press,
  2005).

\bibitem{Tricot2018}
F.~Tricot, D.~H. Phung, M.~Lours, \emph{et~al.}, \enquote{Power stabilization
  of a diode laser with an acousto-optic modulator,}
  {\protect\JournalTitle{Review of Scientific Instruments}} \textbf{89}, 113112
  (2018).

\bibitem{Zhou1996}
W.~L. Zhou, Y.~Mori, T.~Sasaki, and S.~Nakai, \enquote{High-efficiency
  intracavity continuous-wave ultraviolet generation using crystals
  {CsLiB6O10}, β-{BaB2O4} and {LiB3O5},} {\protect\JournalTitle{Optics
  Communications}} \textbf{123}, 583--586 (1996).

\bibitem{Pan2002}
F.~Pan, X.~Wang, G.~Shen, and D.~Shen, \enquote{Cracking mechanism in {CLBO}
  crystals at room temperature,} {\protect\JournalTitle{Journal of Crystal
  Growth}} \textbf{241}, 129--134 (2002).

\bibitem{Yuan2006}
X.~Yuan, G.~Shen, X.~Wang, \emph{et~al.}, \enquote{Growth and characterization
  of large {CLBO} crystals,} {\protect\JournalTitle{Journal of Crystal Growth}}
  \textbf{293}, 97--101 (2006).

\bibitem{Seryotkin2013}
Y.~V. Seryotkin, E.~A. Fomina, and L.~I. Isaenko, \enquote{Humidity effect on
  hydration of {CsLiB6O10} nonlinear optical crystal: {X}-ray diffraction
  study,} {\protect\JournalTitle{Optical Materials}} \textbf{35}, 1646--1651
  (2013).

\bibitem{Morimoto2001}
Y.~Morimoto, S.~Miyazawa, Y.~Kagebayashi, \emph{et~al.},
  \enquote{Water-associated surface degradation of {CsLiB6O10} crystal during
  harmonic generation in the ultraviolet region,}
  {\protect\JournalTitle{Journal of Materials Research}} \textbf{16},
  2082--2090 (2001).

\bibitem{Dubietis2000}
A.~Dubietis, G.~Tamošauskas, A.~Varanavičius, and G.~Valiulis,
  \enquote{Two-photon absorbing properties of ultraviolet phase-matchable
  crystals at 264 and 211 nm,} {\protect\JournalTitle{Applied Optics}}
  \textbf{39}, 2437 (2000).

\bibitem{Isaenko2001}
L.~I. Isaenko, A.~Dragomir, J.~G. McInerney, and D.~N. Nikogosyan,
  \enquote{Anisotropy of two-photon absorption in {BBO} at 264 nm,}
  {\protect\JournalTitle{Optics Communications}} \textbf{198}, 433--438 (2001).

\bibitem{Kondratyuk2002}
N.~Kondratyuk and A.~A. Shagov, \enquote{Nonlinear absorption at 266 nm in
  {BBO} crystal and its influence on frequency conversion,}
  {\protect\JournalTitle{ICONO 2001: Nonlinear Optical Phenomena and Nonlinear
  Dynamics of Optical Systems}} \textbf{4751}, 110--115 (2001).

\bibitem{Divall2005}
M.~Divall, K.~Osvay, G.~Kurdi, \emph{et~al.}, \enquote{Two-photon-absorption of
  frequency converter crystals at 248 nm,} {\protect\JournalTitle{Applied
  Physics B}} \textbf{81}, 1123--1126 (2005).

\bibitem{Kurdi2005}
G.~Kurdi, K.~Osvay, J.~Klebniczki, \emph{et~al.},
  \enquote{Two-photon-absorption of bbo, clbo, kdp and ltb crystals,}
  {\protect\JournalTitle{Advanced Solid-State Photonics}} p. MF18 (2005).

\bibitem{Takahashi2011}
M.~Takahashi, A.~Osada, A.~Dergachev, \emph{et~al.}, \enquote{Improved fourth
  harmonic generation in β-{BaB2O4} by tight elliptical focusing perpendicular
  to walk-off plane,} {\protect\JournalTitle{Journal of Crystal Growth}}
  \textbf{318}, 606--609 (2011).

\bibitem{Bhandari2012}
R.~Bhandari, T.~Taira, A.~Miyamoto, \emph{et~al.}, \enquote{{\textgreater} 3
  {MW} peak power at 266 nm using {Nd}:{YAG}/ {Cr}$^{\textrm{4+}}$:{YAG}
  microchip laser and fluxless-{BBO},} {\protect\JournalTitle{Optical Materials
  Express}} \textbf{2}, 907--913 (2012).

\bibitem{Kumar2015}
S.~C. Kumar, J.~C. Casals, J.~Wei, and M.~Ebrahim-Zadeh, \enquote{High-power,
  high-repetition-rate performance characteristics of
  β-{BaB}$_{\textrm{2}}${O}$_{\textrm{4}}$ for single-pass picosecond
  ultraviolet generation at 266 nm,} {\protect\JournalTitle{Optics Express}}
  \textbf{23}, 28091--28103 (2015).

\bibitem{Yap1998}
Y.~K. Yap, T.~Inoue, H.~Sakai, \emph{et~al.}, \enquote{Long-term operation of
  cslib$_6$o$_{10}$ at elevated crystal temperature,}
  {\protect\JournalTitle{Optics Letters}} \textbf{23}, 34 (1998).

\bibitem{Chang2008}
C.-H. Chang, R.~K. Heilmann, M.~L. Schattenburg, and P.~Glenn, \enquote{Design
  of a double-pass shear mode acousto-optic modulator,}
  {\protect\JournalTitle{Review of Scientific Instruments}} \textbf{79} (2008).

\bibitem{Donley2005}
E.~A. Donley, T.~P. Heavner, F.~Levi, \emph{et~al.}, \enquote{Double-pass
  acousto-optic modulator system,} {\protect\JournalTitle{Review of Scientific
  Instruments}} \textbf{76} (2005).

\end{thebibliography}

\end{document}